# Results Merging in the Patent Domain

VASILEIOS STAMATIS

International Hellenic University

MICHAIL SALAMPASIS

International Hellenic University

In this paper, we test machine learning methods for results merging in patent document retrieval. Specifically, we examine random forest, decision tree, support vector machine (SVR), linear regression, polynomial regression, and deep neural networks (DNNs). We use two different methods for results merging, the multiple models (MM) method and the global model method (GM). Furthermore, we examine whether the ranking of the document's scores is linearly explainable. The CLEF-IP 2011 standard test collection was used in our experiments. The random forest produces the best results in comparison to all other models, and it fits the data better than linear and polynomial approaches.

## 1 INTRODUCTION

Patent documents are distributed in different controlled datasets, patent offices, and other resources that typically must be accessed using different patent search systems and online services e.g. google patents, espacenet, and many more. In some patent search tasks, and for several legal and other reasons, it is crucially important to search in many if possible all relevant resources to increase coverage. To that end, Distributed Information Retrieval (DIR) methods and the Federated Search (FS) approach aim to provide a solution to this problem. DIR as a process can be divided into three different sub-processes. Source representation, source selection, and results merging. The focus of this paper is on the last part. Research has shown that results merging is a very important phase in the DIR process [2] [3] [4] [5], and even if the other sub-processes work satisfactorily if the results merging phase does not operate effectively, the effectiveness of the final results will deteriorate.

Results merging problems were studied by IR research many years ago. Many methods have been developed for results merging but not specifically for the patent industry. The important difference is that patent search many times is recall-oriented since a single missed prior-art for example can cause important economic loss. In terms of Machine Learning (ML), usually linear regression has been used to solve the results merging problem [4]. It is important to examine other ML models than linear regression to assess if they can solve the results merging problem better, especially in the patent industry. Our work presented in this paper addresses this need therefore we propose and test new ML methods for results merging. We used machine learning models to fit the data for results merging and compare the results with other traditional approaches. Second, we test our models in different environments and we investigate the relation of document scores in a ranked list of documents.

## 2 RELATED WORK

Patent retrieval is a subfield of information retrieval. Whilst information retrieval has progressed a lot in terms of research and development, patent retrieval is a more traditional and complex, therefore more challenging area [6]. The results merging problem was studied as a general DIR problem and not in the specific context of the patent domain. The work that has been done by Voorhes, Gupta, and Johnson-laird in [7] was one of the first to conduct experiments in results merging. After that many algorithms appeared in research. Taylor, Radlinski, and Shokouhi in [8] published a patent for results merging using a machine learning process. Another patent published about merging results lists according to scores assigned to the lists and the documents by Mao et al. in [9].

The collection inference retrieval network CORI [3] uses weighted score merging and it is considered as a very stable method performing very good or state-of-the-art results in many experiments. CORI is a linear combination of the source selection score and the score of the document returned by the collection and uses a simple heuristic formula to normalize the collection-specific scores and transform them into global comparable scores able to be used to produce a single merged list.

Another effective method is the semi-supervised learning algorithm (SSL) [5]. This method uses linear regression to estimate using local collection-specific scores the global comparable scores. For that purpose, their algorithm is based on the common documents returned from each remote collection and a centralized sample index consisting of sample documents from all the different collections.

SAFE (sample-agglomerate fitting estimate) is a more recent algorithm that is designed to function with minimum cooperation between the broker and the collections [10]. SAFE is based on the principle that for a given query, the results of the sampled documents are a sub-ranking of the original collection, so curve fitting to this sub-ranking can be used to estimate the original scores.

Lee et al. in [11] proposed an optimization framework for results merging. They used λ-merge method [12] and they extended it for implementing results merging by adding the extra component of the vertical quality (resource quality) in this framework.

## 3 METHODOLOGY

In the experiments that we report in this paper, we used the CLEF-IP 2011 standard test. To run the experiments in a federated environment, we create an artificial federated environment using the intellectually assigned patent classification codes (IPC/CPC). Similarly to the work done by Salampasis, Paltoglou, and Giahanou [17], we split the collection based on the IPC codes at level 3 (subclass). The IPC/CPC system is an internationally accepted standard taxonomy for classifying, sorting, and organizing patent documents [18]. The split 3 has 632 different IPC codes so this results in 632 indexes or resources that they can be searched simultaneously. To create the indices we used anserini, a toolkit for information retrieval research which is based on Lucene [19]

There were 3973 topics at the CLEF-IP 2011 campaign which were in three languages (English, German, French). We use the first 300 English topics to create the queries. Each query was no more than 1000 words coming from the title, abstract, first 500 words of the description, and claims.

We test the algorithms both in cooperative settings where documents' scores and collection statistics like term frequencies etc. from remote collections are available and uncooperative settings which is the case most of the time in real world, and the returned documents are only ranked lists of documents without scores. We solve the lack of relevancy scores to the documents by assigning artificial scores in the documents according to their ranking and then multiplying the documents artificial score with the score the resource has taken at the source selection phase. We assign 0.6 to the first document and descending by even increments, we assign 0.4 to the last. We chose these scores as they have been shown to work well in the CORI algorithm [22]. Then we multiplied all these scores with the respective source selection score that was assigned to the specific source.

The algorithm we used for all source selection processes is CORI. Another parameter that is considered is the total number of remote collections to route the query. In other words, how many of the 632 available collections will be requested to return their search result to be considered in the merging phase. After several tests, we chose 20.

### 3.1 Results Merging

For merging the results the algorithms take into account the overlapping documents between the retrieved documents from each collection and the documents retrieved from the centralized index. The centralized index is created as follows. We used query-based sampling [21], a method used for creating representations of collections that can be used to approximate the statistics of federated collections for uncooperative environments where statistics of collections are not available. We created representation sets of all the collections i.e. 632 and we created a centralized index consisting of all the sampling documents. Each representation set consisting of around 300 documents. We used all the sampling documents from all the collections and we created a centralized index.

The overlapping documents are the common documents between the returned documents from the collections and the returned documents from the centralized index. The overlapping documents are used to train the models to convert local collection-specific scores to the global scores that the centralized index would have assigned. The scores from the centralized index are comparable as the samples of all the collections co-exist in it. In other words, the models are trained to convert local collection-specific and non-comparable scores to global comparable scores so the final merging can be finally applied.

## 3.2 Machine Learning Methods

We used two different methods for merging the results. In the first set of experiments, we used the overlapping documents between the collections and the centralized index and we train one model for each collection. Each model was then used to calculate the global relevance scores for the rest of the non-common documents returned by the collection.

In the second set of experiments, we used global machine learning model for all the collections per query. These Global models (GM) take as input the returned documents' scores from all the collections. For example, if we choose to submit the query to ten sources, the algorithm's inputs are ten and they are the documents' scores returned from the sources if the document returned by the source, otherwise it is zero. The main motivation to implement these models in that way is because when a document is returned by more than one collection, this information will be taken into consideration.

## 4 RESULTS

We ran the experiment two times in a cooperative and uncooperative environment as already mentioned. First, we assume a cooperative environment where document scores returned. Second, we assign artificial scores in conjunction with the source selection score. At each experiment, we compare the global models, multiple models, and between them. Also, we implement CORI and SSL two state-of-the-art methods to examine the results merging efficiency.

### 4.1 Cooperative environment

We first assume a cooperative environment so we used the scores as returned from the collections. For the multiple models (MM), Table 1 summarizes the results. The best performance, in general, comes from the multiple models. The random forest gave the best results. We got a similar performance from the decision tree as well. Both the random forest and decision tree also overcame the two state-of-the-art algorithms SSL and CORI at all three metrics. Furthermore, SSL performs better than polynomial regression. We created a polynomial regression based on SSL. We used the linear regression model and we add the polynomial features $x^2$ and $x^3$. This suggests that linear mapping to the ranking scores is better than polynomial mapping.

For the global models, the deep neural network was created with 4 hidden layers with [632,300,150,50] neurons at each hidden layer respectively. The activation function was "relu" and we used Adam optimizer with a learning rate of 0.01. The loss function we used was the mean squared error. The best performance from global models in terms of MAP was from the deep neural network. Random forest gave the best PRESS and RECALL scores. Except for the support vector machine all the global models gave better PRESS and RECALL results than CORI. This is important because the patent industry is recall-oriented as missing patent documents can have a huge economic impact. SSL performed better than all global models. SVR performed well in the MM, here in the GM seems to perform quite badly compared to other models. This might have to do with the function of SVR when it split clusters of data in conjunction with the big differences in document scores that different sources might return. Additionally, SSL performed better than CORI. The centralized approach gave the best MAP than all DIR methods.

### 4.2 Uncooperative Environment

For the multiple models, we used random forest, SVR, decision tree, and polynomial regression. The best performance algorithm at all three metrics was again the MM random forest following by the MM decision tree. From the multiple models, random forest, decision tree, and SVR overcame both CORI and SSL in PRESS and RECALL score. SSL performed better than both polynomial regressions so again the linear mapping is better than polynomial mapping.

For the global models, the best performing algorithm here is the random forest in terms of MAP and the linear regression in terms of PRESS and RECALL. The deep neural network was very close to the previous two. Both CORI and SSL performed better than all global models in terms of MAP but random forest, linear regression, decision tree, and the deep neural network has greater PRESS and RECALL scores.

Looking at CORI and SSL it can be observed that CORI overcame SSL at MAP. This is an interesting finding as it suggests that CORI is more robust than SSL in terms of MAP when assigning local scores. In the cooperative environment, SSL performs better than CORI and this is consistent with other findings reported in the literature [5]. Comparing all the models, MM random forest is the best performing model following by the decision tree. Also, random forest performed better than the centralized approach. Table 1 summarizes the results.

Table 1: Scores of models in both environments

|  | Cooperative Environment | | | Uncooperative environment | | |
| --- | --- | --- | --- | --- | --- | --- |
|  | MAP @100 | PRESS @100 | RECALL @100 | MAP @100 | PRESS @100 | RECALL @100 |
| MM Random Forest | 0.0777 | 0.3358 | 0.3468 | 0.0837 | 0.2674 | 0.2738 |
| MM SVR | 0.0612 | 0.2679 | 0.2779 | 0.0709 | 0.2348 | 0.2413 |
| MM Decision Tree | 0.0745 | 0.3283 | 0.3391 | 0.0774 | 0.2672 | 0.2740 |
| MM Polynomial $x^2$ | 0.0454 | 0.2036 | 0.2105 | 0.0437 | 0.1182 | 0.1200 |
| MM Polynomial $x^3$ | 0.0299 | 0.1243 | 0.1287 | 0.0440 | 0.1252 | 0.1275 |
| GM Random Forest | 0.0465 | 0.2406 | 0.2483 | 0.0460 | 0.2434 | 0.2513 |
| GM SVR | 0.0218 | 0.0788 | 0.0820 | 0.0217 | 0.0846 | 0.0882 |
| GM Decision Tree | 0.0414 | 0.2315 | 0.2391 | 0.0420 | 0.2318 | 0.2394 |
| GM Linear Regression | 0.0517 | 0.2401 | 0.2474 | 0.0436 | 0.2444 | 0.2523 |
| GM DNN | 0.0693 | 0.2353 | 0.2416 | 0.0412 | 0.2434 | 0.2510 |
| CORI | 0.0650 | 0.2102 | 0.2161 | 0.0714 | 0.1940 | 0.1969 |
| SSL | 0.0725 | 0.2464 | 0.2528 | 0.0623 | 0.2168 | 0.2219 |
| Centralized | 0.0793 | 0.2592 | 0.2660 | 0.0793 | 0.2592 | 0.2660 |

In summary, MM random forest was the best model and produced significant improvement compared to CORI and SSL in both environments. This proves its robustness and also that it can fit the documents score in a ranking better than linear and polynomial functions. Finally, the new method for assigning local scores in uncooperative environments seems to be promising as in some cases like the GM random forest we got higher results than the respective in the cooperative environment.

## 5 CONCLUSION

In this paper, ML models for results merging are proposed and tested. Also, a new method for assigning scores to documents in an uncooperative environment is presented and the document scores in a ranking are investigated. We tested random forest, decision tree, linear regression, polynomial regression, SVR, DNN in multiple models, and global model conditions, and the multiple model random forest was the best model which increased the results compared to traditional approaches up to +60%. Random forest was the best at all different environments and this shed its robustness. Also, it managed to overcome the centralized approach and this suggests that random forest could be a better option especially for the patent industry where missing a single patent can cause important economic loss.

In uncooperative environments assign artificial scores and multiplying with the source selection score increased the results and seems to be a better way to assign document scores. For the document scores ranking, linear regression is better than polynomial regression but the random forest is even better than both, so linear mapping to the scores is not the best option.

For future work, we plan to further investigate the models and try a different combination of parameters. Also, we plan to create reusable models that can be trained and reused. We plan to train our models on a larger dataset and run larger experiments.

## ACKNOWLEDGMENTS

This work has received funding from the European Union's Horizon 2020 research and innovation programme under the Marie Skłodowska-Curie grant agreement No: 860721